\documentclass[twocolumn,showpacs]{revtex4}
\usepackage{graphicx}
\graphicspath{{pict/}{}}

\usepackage{bm}

\newcounter{Fig}

\newcommand{\be}{\begin{equation}}
\newcommand{\ee}{\end{equation}}

\begin{document}
\title{Boundary-induced Anderson localization in photonic lattices}
\author{Mario I. Molina}
\affiliation{Departamento de F\'{\i}sica, Facultad de Ciencias,
Universidad de Chile, Casilla 653, Santiago, Chile\\
Center for Optics and Photonics, Universidad de Concepci\'{o}n, Casilla 4016, Concepci\'{o}n, Chile}

\begin{abstract}
We analyze numerically localization of light in linear square waveguide
arrays restricted in one dimension (``ribbons''), whose boundaries
are disordered in propagation constant and/or coupling.
We find that the disordered boundary induces a localization
tendency in the bulk even for relatively wide ribbons.
\end{abstract}

\pacs{42.65.Wi,42.65.Tg,42.81.Qb,05.45.Yv}

\maketitle

The phenomenon of Anderson localization (AL) in disordered system constitutes one of the staples of modern condensed-matter physics. Proposed originally for electrons and one-particle excitations in solids\cite{solid1,solid2,solid3,solid4}, it was soon extended to many other fields such as acustics\cite{acustic1,acustic2,acustic3}, Bose-Einstein
condensates\cite{BE} and optics\cite{optic1,optic2,optic3,optic4,optic5}. It still continues to surprise us even now, more than 50 years since its discovery\cite{anderson}.

The AL effect is ultimately due to wave interference between multiple-scattering paths, which makes optical systems an ideal setting for exploring the AL phenomenon, without encumbering effects such as electron-electron interactions in solids, or other many-body effects. Thus far, most research carried on in optical disordered systems has focussed mainly on bulk disorder. However, interesting localization effects due to surface disorder have  been found in single-and-multimode waveguides. For instance, localization of waves due to multiple-scattering by surface roughness has been predicted for a single-mode waveguide and for a thin film\cite{frei1,frei3,frei2}.

In the same spirit, we consider in this work the case of a two-dimensional photonic lattice with boundary disorder. As we show below, for finite width ribbons, boundary disorder does induce AL in the bulk.

Let us consider a two-dimensional square linear  $N\times M$ waveguide array with a finite extension in one dimension (Fig.1), i.e., a ``ribbon'' ($M \ll N$). In the framework of the coupled-modes theory, the electric field  $E({\bf r},z)$ propagating along the waveguides can be presented as a superposition of the waveguide modes, $E({\bf r},z)=\sum_{\bf n} E_{\bf n}(z) \phi ({\bf r}-{\bf n})$, where
${\bf r}=(x,y)$, $E_{\bf n}$ is the amplitude (in units of (W)$^{1/2}$) of the single guide mode $\phi(\bf r)$ centered
on site with the lattice number ${\bf n}=(n_{1}, n_{2})$. The evolution
equations for the modal amplitudes $E_{\bf n}$ are
\begin{equation}
i {d E_{\bf n}(z)\over{d z}} + \epsilon_{\bf n} E_{\bf n}(z) + \sum_{\bf m} V_{{\bf n},{\bf m}} E_{\bf m}(z) = 0,\label{eq:1}
\end{equation}
where ${\bf n}$ denotes the position of the guide center, $z$ is the longitudinal distance (in meters), $V_{{\bf n},{\bf m}}$ is the coupling between nearest-neighbor guides (in units of 1/m) and $\epsilon_{\bf n}$ is the propagation constant of guide with center at ${\bf n}$.

{\em Stationary states}.  The stationary modes of the system are found by posing a solution of the form
$E_{\bf n}(z) = C_{\bf n} \exp(i \beta z)$ in Eq(\ref{eq:1}), which leads to
\begin{equation}
-\beta C_{\bf n} + \epsilon_{\bf n} C_{\bf n} + \sum_{\bf m} V_{{\bf n},{\bf m}} C_{\bf m} = 0.\label{eq:2}
\end{equation}
To ascertain the localization properties of a given mode of the system, we use the inverse participation ratio
(IPR), defined by $IPR = \sum_{\bf n} |C_{\bf n}|^{4}/\sum_{\bf n} |C_{\bf n}|^{2}$. For completely localized modes, $IPR=1$, while for completely delocalized modes, $IPR=1/N$, where $N$ is the number of sites of the lattice. We will average the IPR over all states of the system and also over a number of random realizations, to obtain the average IPR which reflects the global localization tendency of the system.
\begin{figure}[t]
\includegraphics[height=5.5cm, angle=0]{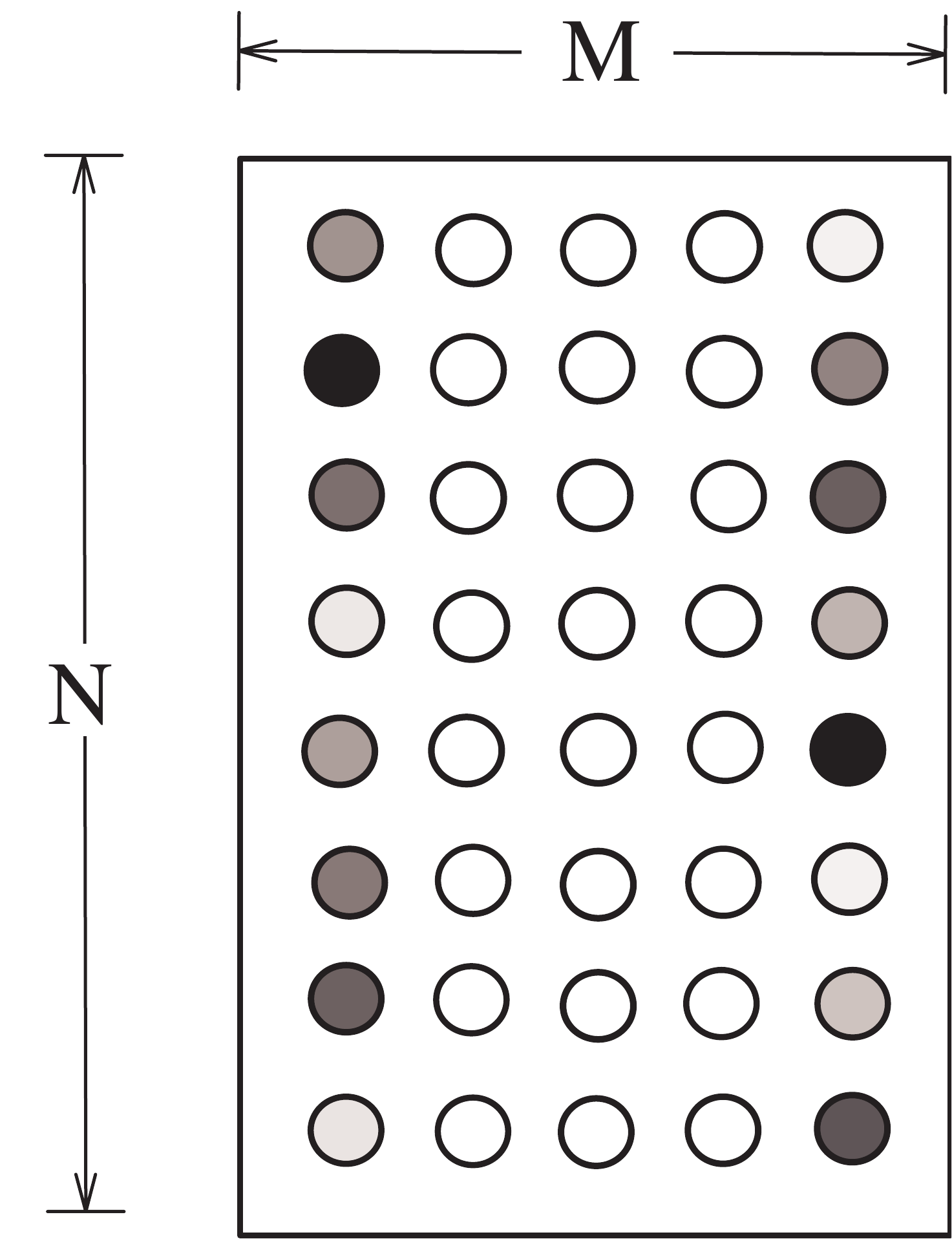}
\caption{Example of optical lattice ribbon ($M\ll N$) with disordered boundary. The propagation constants along the long boundary have random values.}
\label{fig1}
\end{figure}
We start with the case where the propagation constants $\epsilon_{\bf n}$ along the boundary of the ribbon (gray sites on Fig.\ref{fig1}) take on random values stemming from a uniform distribution of width $w$: $\epsilon_{\bf n}\in [-w,w]$, while coupling between nearest-neighbor guides is identical ($\equiv 1$). Results for the
average IPR are shown on Fig.\ref{fig2}(a), for several disorder widths, including the case of no disorder $w=0$, which is necessary for comparison, since we are dealing with finite systems.
\begin{figure}
\includegraphics[height=4.5cm]{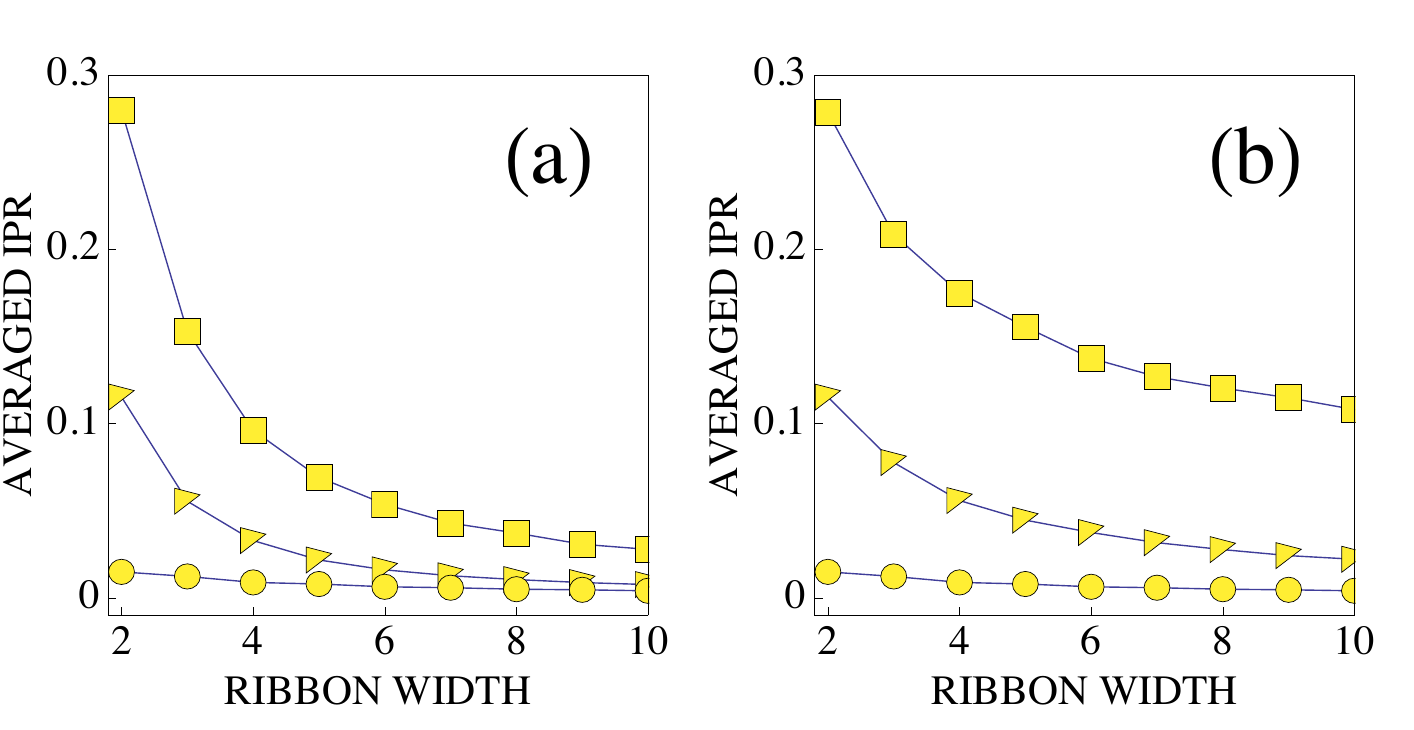}\\
\vspace{-0.25cm}
\includegraphics[height=4.5cm]{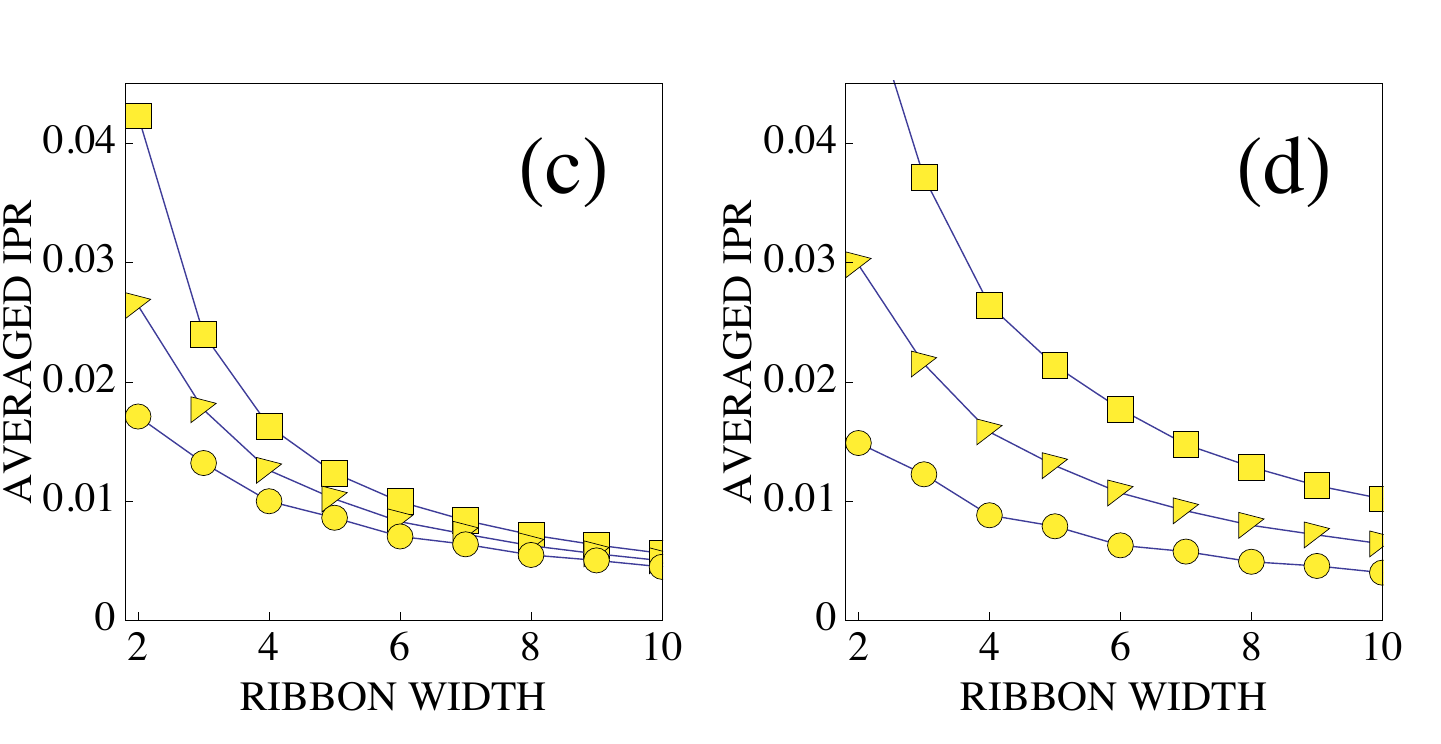}
\caption{State-and-realization average inverse participation
ratio  vs ribbon width for (a)-(b) {\it disorder in waveguide propagation constant},
and (c)-(d) {\it disorder in waveguide coupling}, 
for several disorder widths: $w=0$(circles), $w=1$ (triangles) and
$w=2$ (squares). Left (right) column corresponds to surface (bulk) disorder.}
\label{fig2}
\end{figure}
Figure \ref{fig2}(b) shows the corresponding results for a ribbon with {\em all} its propagation constants random. As expected,  the IPR is smaller for the surface disorder case than  for the bulk disorder case. But we can also see that for boundary disorder, some localization tendency is still appreciable for intermediate widths, e.g.,  $M=4, 5$.
\begin{figure}
\includegraphics[height=7cm]{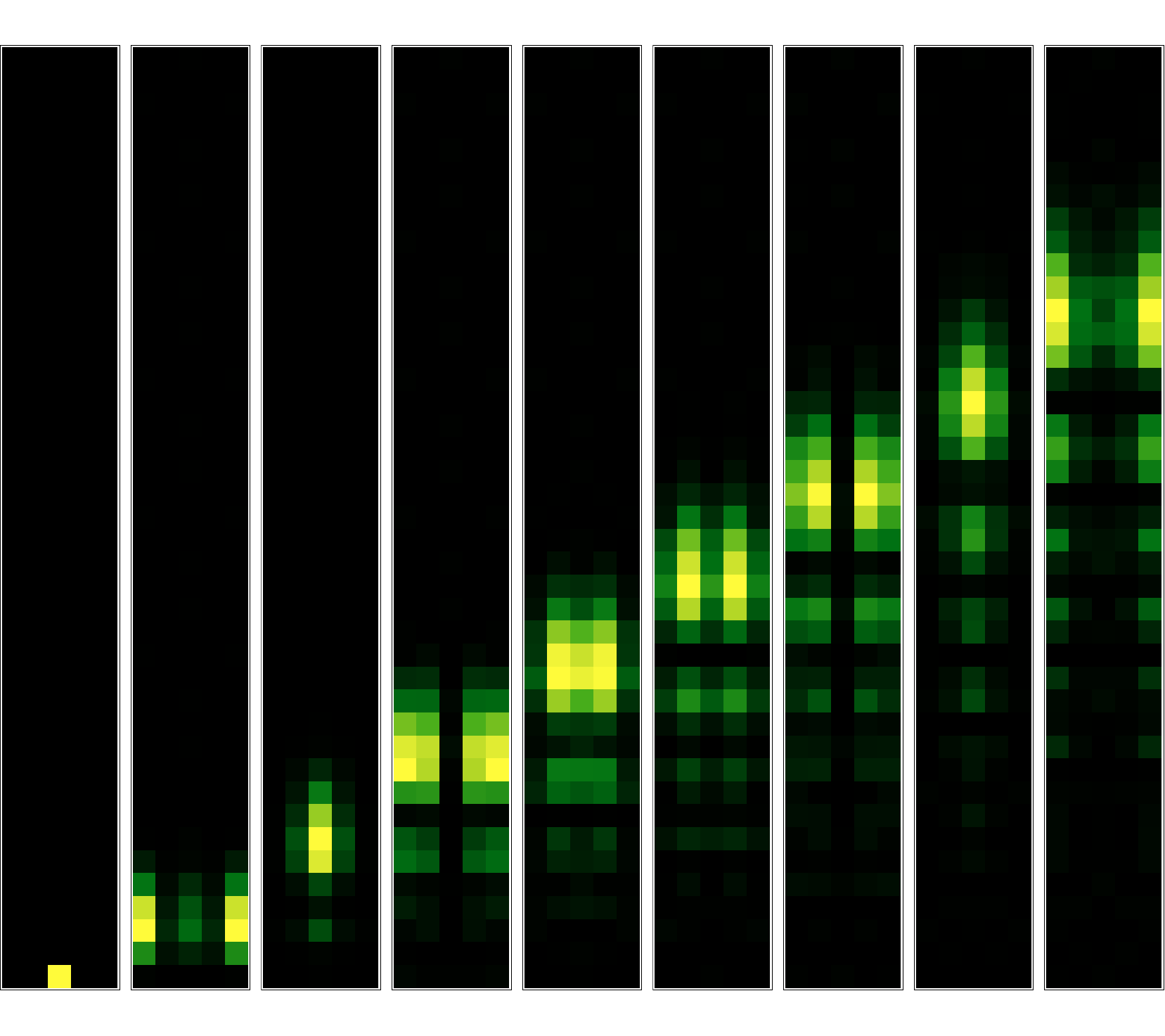}
\caption{Dynamical evolution of an initially localized input
beam at the boundary of a 41$\times$5 completely ordered photonic ribbon. From left to
right we show snapshots at $V z=0,2,4,6,8,10,12,14,16$}
\label{fig4}
\end{figure}
For the case where $\epsilon_{n}\equiv 0$, but the couplings along the boundary of the ribbon (between shaded sites on Fig.\ref{fig1}) take on random values from a uniform distribution: $V_{{\bf n},{\bf m}}=1 + \delta V$, where $\delta V\in [-w/4,w/4]$, Fig.\ref{fig2} shows qualitatively similar results as for the disordered propagation constants case, although the localization effect is smaller in this case.
The above results suggest the idea of a disorder ``memory'' in intermediate width boundary-disordered ribbons.

\begin{figure}[h]
\includegraphics[height=7cm]{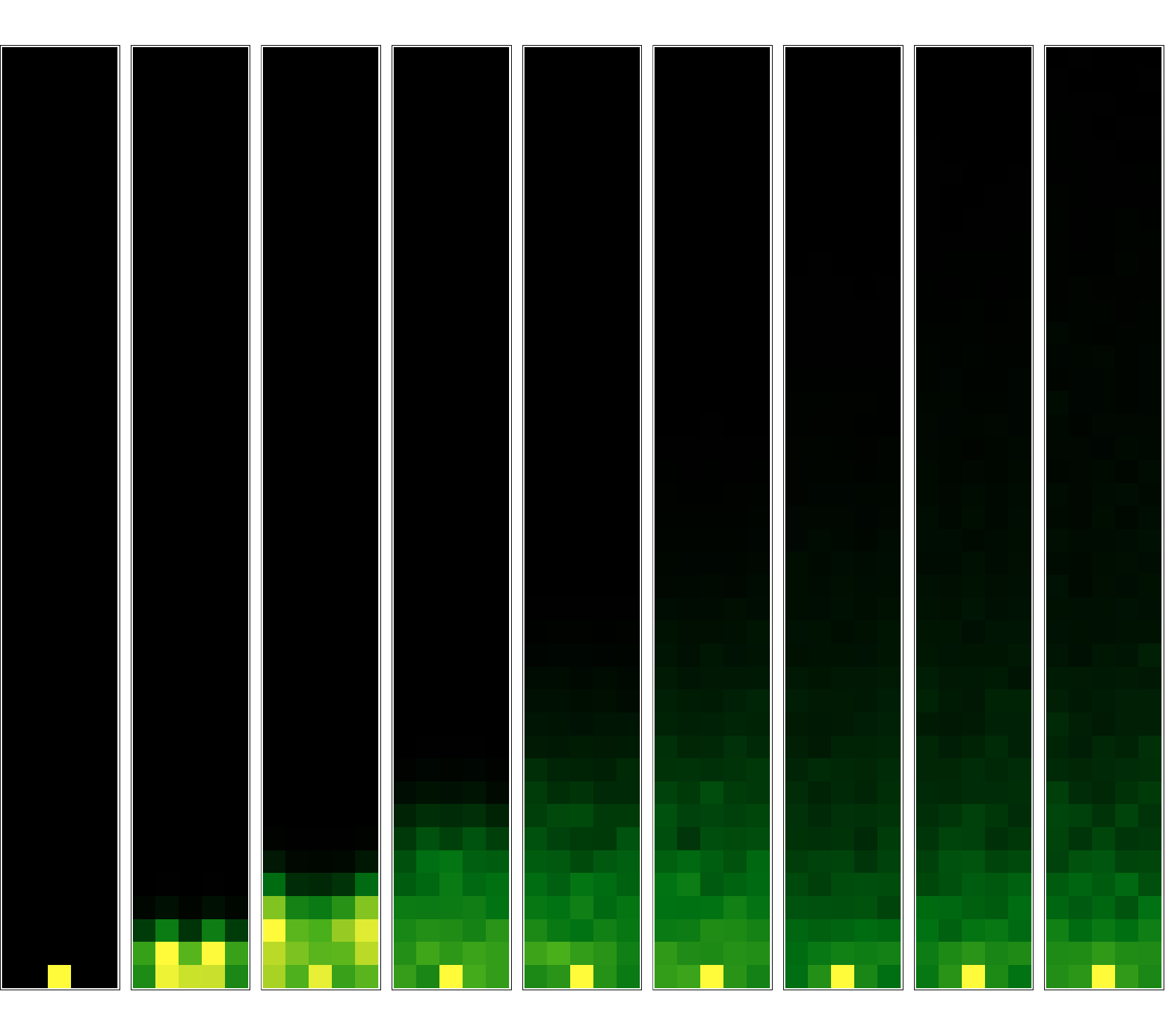}
\caption{Average over 100 realizations of dynamical evolution of an initially localized input
beam at the boundary of a 41$\times$5 bulk-disordered ($\epsilon_{\bf n}\in [-1,1]$ for all ${\bf n}$) photonic ribbon. From left to
right we show snapshots at $V z=0,1,2,4,8,16,32,64,128$.}
\label{fig5}
\end{figure}

{\em Dynamics}.\ We examine now the evolution (Eq.\ref{eq:1}) of an initially localized input beam launched at the middle site belonging to one of the not-disordered boundaries of the ribbon, $(0,M/2)$.  The long boundaries are disordered, with random propagation constants $\epsilon_{\bf n}$. The idea is to examine the evolution of the spatial  optical power distribution, and compare it to the cases with no disorder ($\epsilon_{\bf n}=\mbox{constant}$) and to the case with bulk disorder ($\epsilon_{\bf n}=\mbox{random}$ for all sites). Since the problem is linear, we can express $E_{\bf n}(z)$ as a superposition of the stationary modes
of the system. For localized initial conditions $E_{\bf n}(0)=\delta_{{\bf n},{\bf n_{0}}}$, with ${\bf n_{0}}=(0, M/2)$, the solution of Eq.(\ref{eq:1}) is $E_{\bf n}(z)= \sum_{l}\ C_{\bf n_{0}}^{(l)*} C_{\bf n}^{(l)} e^{-i z \beta^{(l)}}$ where $C_{\bf n}^{(l)}$ and $\beta^{(l)}$ are the l-th  mode and its associated propagation constant of the eigenvalue Eq.(\ref{eq:2}).
For the completely ordered array (Fig.\ref{fig4}), the beam spreads out, away from the boundary, and most of its power is carried in the frontal lobes. 
In the case of bulk disorder (Fig.\ref{fig5}) we see the usual Anderson localization phenomena with the optical power remaining  confined within a small region around the excitation point.
For the case of boundary disorder , however, something very interesting happens: After a short transient longitudinal distance, the extent of which depends on both, the width of the disorder $w$ and the relative width of the ribbon $M$, the system tends to localize a finite fraction of its power at the initial launching site, while the
rest of the
\begin{figure}[t]
\includegraphics[height=7.25cm]{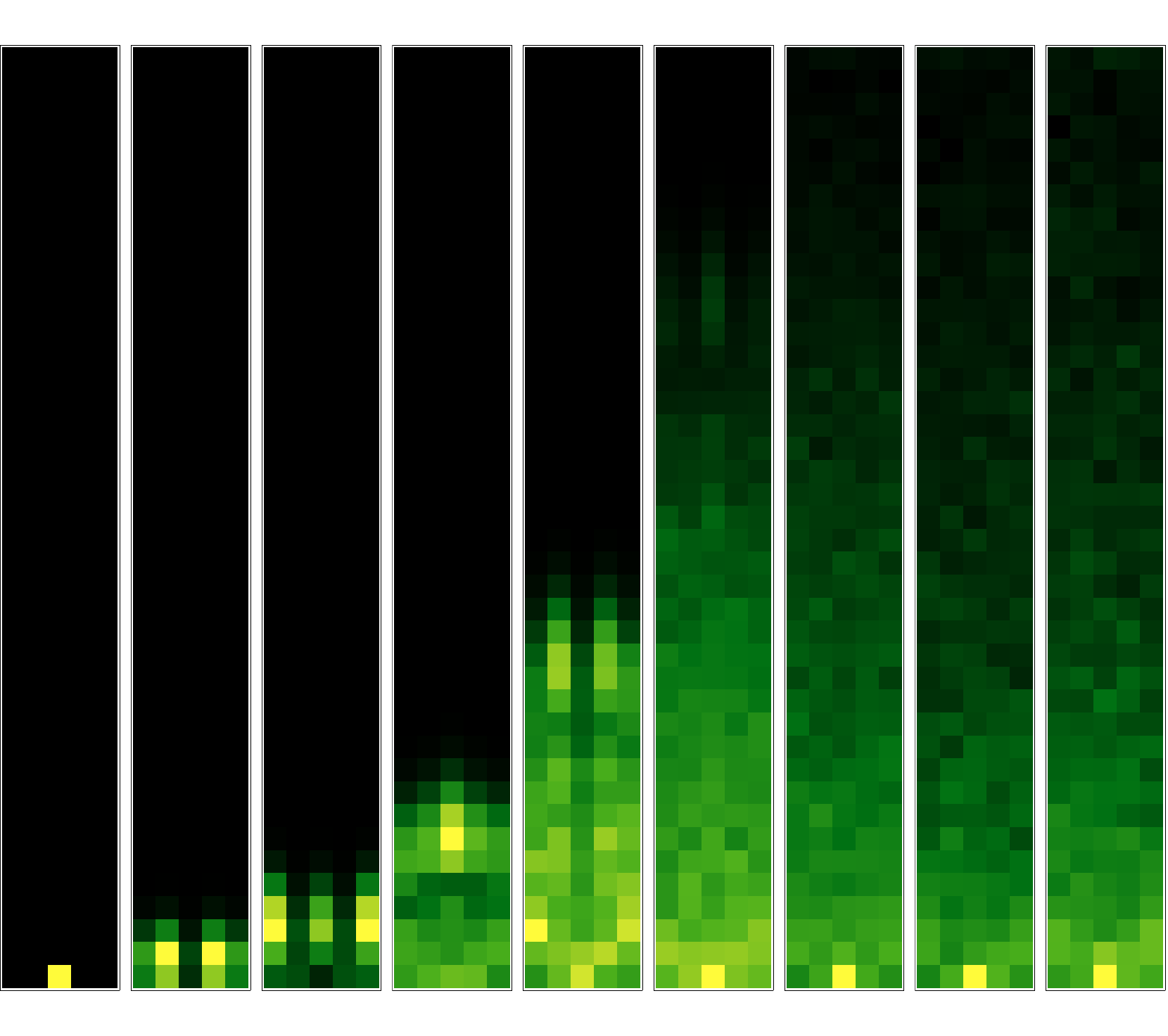}
\caption{Average over 100 realizations of the dynamical evolution of an initially localized input beam at the boundary of a 41$\times$5 boundary-disordered ($\epsilon_{\bf n}\in [-1,1]$ along the long boundaries) photonic ribbon. From left to
right we show snapshots at $V z=0,1,2,4,8,16,32,64,128$}
\label{fig6}
\end{figure}
optical power gets eventually Anderson-localized (Figs.\ref{fig6} and \ref{fig7}). The effect is stronger for narrow ribbons (e.g., $5$ sites), but has been verified for wider ribbons as well, albeit,
with a greater localization evolution distance (``time''). This is only natural, since for wider ribbons it takes longer to achieve enough multiple scattering for localization effects. For the case when the disordered lies only at the narrow edge where the beam is launched (narrow lower edge in Fig.1), the above localization effect ceases, and we only obtain partial localization at the input site, plus free propagation of the remaining fraction (not shown). Thus, it seems that multiple wave scattering between at least two disordered boundaries is needed to effect AL in these kind of photonic lattices.
\begin{figure}[h]
\includegraphics[height=5cm]{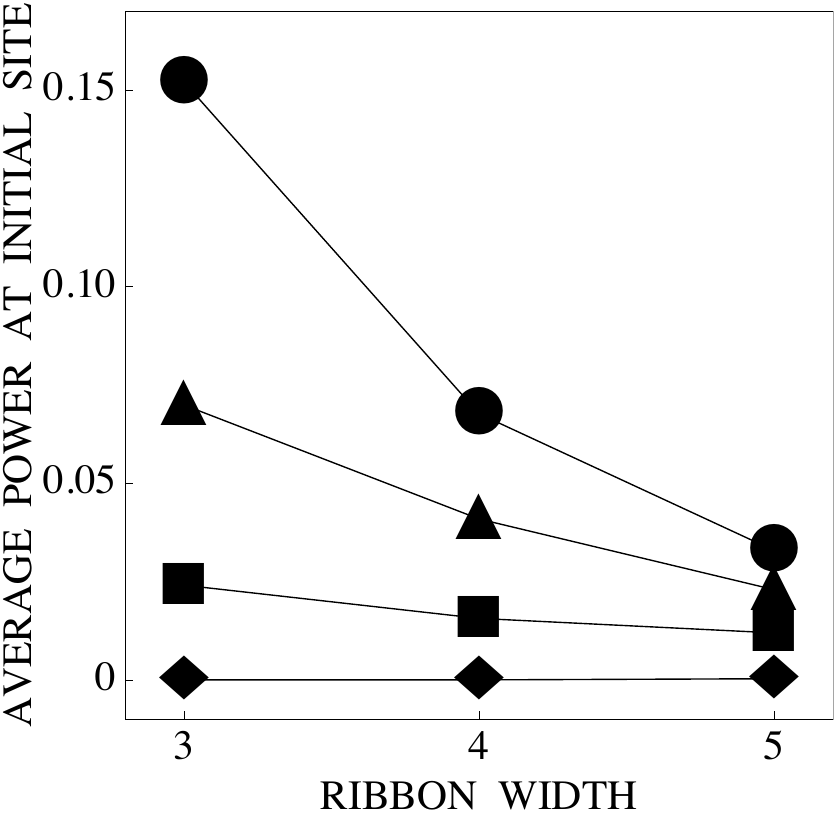}
\caption{Time and realization (100) average power remaining at
initial guide vs ribbon width, for disorder width $w=0$ (rhombi), $w=0.5$ (squares), $w=1$ (triangles) and $w=2$ (circles).}
\label{fig7}
\end{figure}

Therefore, the presence of boundary disorder does effect Anderson localization on the whole system. This effect is reminiscent of localization in a
single-mode waveguide with rough edges\cite{frei3}, and suppression of thermal conductivity in thin graphene nanoribbons with rough edges\cite{ysk} .

In summary, we have analyzed numerically localization effects on a two-dimensional photonic lattice of finite width with disordered boundaries,
by computing the average inverse participation ratio of the stationary modes, and by dynamical evolution of an initially localized input beam. We conclude that a disordered boundary does induce AL effects on the whole bulk.

The author is grateful to Y.S. Kivshar for useful discussions.
The author acknowledges support from FONDECYT grants 1080374, 1070897, and
Programa de Financiamiento Basal de CONICYT (grant FB0824/2008).

\end{document}